\documentclass[conference]{IEEEtran}

\IEEEoverridecommandlockouts

\usepackage[capitalise,noabbrev]{cleveref}
\usepackage[group-separator={,},group-minimum-digits=4]{siunitx}
\usepackage{dirtytalk}
\usepackage{listings}
\usepackage[hyphens]{url}

\newcommand{\os}{operating system}
\newcommand{\mem}{memory}
\newcommand{\all}{allocator}
\newcommand{\memall}{\mem\ \all}
\newcommand{\oss}{open-source software}
\newcommand{\krn}{kernel}

\makeatletter
\def\lst@makecaption{
  \def\@captype{table}
  \@makecaption
}
\makeatother

\begin{document}

\title{There Ain't No Such Thing as a\\Free Custom Memory Allocator}

\makeatletter
\newcommand{\linebreakand}{%
  \end{@IEEEauthorhalign}
  \hfill\mbox{}\par
  \mbox{}\hfill\begin{@IEEEauthorhalign}
}
\makeatother

\author{
  \IEEEauthorblockN{Gunnar Kudrjavets}%
  \IEEEauthorblockA{\textit{University of Groningen}\\
   9712 CP Groningen, Netherlands\\
    g.kudrjavets@rug.nl}
  \and
  \IEEEauthorblockN{Jeff Thomas}
  \IEEEauthorblockA{\textit{Meta Platforms, Inc.} \\
    Menlo Park, CA 94025, USA \\
   jeffdthomas@fb.com}
  \and
  \IEEEauthorblockN{Aditya Kumar}
  \IEEEauthorblockA{\textit{Snap, Inc.}\\
    Santa Monica, CA 90405, USA \\
    adityak@snap.com}
  \linebreakand %
  \IEEEauthorblockN{Nachiappan Nagappan}
  \IEEEauthorblockA{\textit{Meta Platforms, Inc.} \\
    Menlo Park, CA 94025, USA \\
    nnachi@fb.com}
  \and
  \IEEEauthorblockN{Ayushi Rastogi}
  \IEEEauthorblockA{\textit{University of Groningen}\\
    9712 CP Groningen, Netherlands\\
    a.rastogi@rug.nl}
}

\maketitle

\begin{abstract}
Using custom \memall s is an efficient performance optimization technique.
However, dependency on a custom \all\ can introduce several maintenance-related issues.
We present lessons learned from the industry and provide critical guidance for using custom \memall s and enumerate various challenges associated with integrating them.
These recommendations are based on years of experience incorporating custom \all s into different industrial software projects.
\end{abstract}


\section{Introduction}

A \memall\ is responsible for handling \mem\ management requests coming from an application.
The typical operations that an application can request an \all\ to perform are related to allocating and releasing a specific amount of \mem.
The entirety of \mem\ available to the \os\ is managed by a \krn\ \memall.
A typical application runs in a user mode.
In this paper, we focus on user mode \memall s.

A \emph{default \all}, such as one from \textsc{GNU C} {L}ibrary~\cite{gnuc}, is typically provided by an \os\ or a runtime library of a specific programming language.
One approach to improve the application's performance is to \emph{replace the default \memall\ with a custom one}.
Various case studies~\cite{evans_2006,evans_scalable_2011,berger,durner} document the benefits of using custom \memall s in the industry.
However, the maintenance issues associated with integrating a custom \memall\ with the application's code base have not been studied.
Based on our experience with integrating and using custom \memall s for several years, we present a structured enumeration of various issues associated with maintaining custom \memall s.

\section{Background}

The performance characteristics of an application are important criteria for its success.
Improving performance is also one of the main reasons to use custom \memall s.
Applications like database engines, stock trading systems, or web browsers use custom \memall s~\cite{berger,durner}.
A variety of \all s exist:
 \emph{jemalloc}~\cite{evans_2015},
 \emph{mimalloc}~\cite{leijen2019mimalloc},
 \emph{rpmalloc}~\cite{jansson_rpmalloc},
 \emph{snmalloc}~\cite{snmalloc}, and
 \emph{tcmalloc}~\cite{lee_2014}.
We observe that both industry and \oss\ projects tend to use either \emph{jemalloc} (currently mainly developed by Facebook) or \emph{tcmalloc} (developed by Google).
The \emph{mimalloc} project developed by Microsoft is another industrial-strength \all.

Some \all s were developed for the sole purpose of increasing the performance of a specific application.
For example, the introduction of \emph{PartitionAlloc} into {C}hromium resulted in up to \num{22}\% \mem\ savings on Windows~\cite{partition_alloc}.

Using custom \memall s has multiple benefits:

\begin{itemize}
	\item \textbf{Ability to tune \all's behavior to meet the needs of a specific application}.
	Default \all s are optimized using the one-size-fits-all approach to work at an equally acceptable level for \emph{all the applications}.
    With custom \all s, such as \emph{jemalloc}, several configuration options can be used to optimize the \all 's performance both during the initialization or during the runtime.

	\item \textbf{Improved debugging and tracing facilities}.
	Memory-related defects compose a sizeable portion of post-release defects.
	For example, \num{70}\% of post-release security issues Microsoft has to fix annually are related to \mem\ management~\cite{matt_trends_2019}.
	Typical \mem -related problems are double-frees, heap corruptions, and \mem\ leaks.
	Custom \all s provide an enhanced debugging experience and help engineers fix the defects faster.
	Using an \all\ with improved debugging capabilities is an efficient way to find defects that the default \all\ cannot detect.

	\item \textbf{Access to complete source code}.
	In closed-source \os s, engineers do not have access to the \all 's source code.
	That makes debugging complex issues challenging.
	Using an open-source \all\ introduces several benefits.
	Engineers can add new features or remove unnecessary code to further optimize applications performance.
	Availability of source code also enables engineers to have an improved debugging experience.
\end{itemize}

We have observed that the benefits provided by custom \all s come with the significant cost related to maintenance of the \all's code base alongside the application's code base.
Existing research~\cite{evans_2006,evans_scalable_2011,berger,durner} highlight the benefits, but they do not document the drawbacks and potential problems related to using custom \all\ code.
Using a custom \all\ does not guarantee that the application’s performance will improve.
The behavior of a custom \all\ on the same platform varies significantly~\cite{patel_2017}.

In practice, replacing various components with custom versions is a standard performance optimization technique.
For example, engineers can use a different library to parse \textsc{JSON},
or use an implementation of data structures that are different from those provided by the language runtime.
With those components, engineers typically explicitly call one function instead of another.
Replacing the \memall\ has one crucial difference---\emph{it impacts all the code interacting with the \all\ that is loaded and executed in the same process address space}.

One way to replace a default \all\ is to provide custom implementations for a fixed set of functions that relate to \mem\ management.
Their interface does not change but the implementation does.
Typical functions that \memall s intercept and replace~\cite{posix} are shown in Listing~\ref{code:memfuncs}.

\lstset{language=C,breaklines=true,frame=single,basicstyle=\ttfamily,caption={\textsc{POSIX} compliant \mem\ allocation functions.},label=code:memfuncs}
\begin{lstlisting}
void free(void* ptr);
void* malloc(size_t size);
void* calloc(size_t num, size_t size);
void* realloc(void* ptr, size_t size);
\end{lstlisting}

Replacing \all s is a risky proposition.
Application code, all of its dependencies, and system libraries will have the component responsible for \mem\ management replaced underneath them.
There is no indication that the switch has happened.
An issue related to \mem\ management (either exposed or introduced by a custom \all) has catastrophic consequences for an entire application.
Corrupting a single location in the heap of the current process means that the application will be in an inconsistent state, and its future behavior is undetermined.

\section{Maintenance challenges}

\subsection{Source code management}

\subsubsection{Build support}
\label{subsec:build}

Custom \memall s are mostly developed as \oss.
Various build systems such as Ant, Babel, Buck, {CM}ake, and Ninja exist.
Authors of the custom \all\ support only a limited set of build systems.
However, an application that intends to use a custom \all\ may utilize a build system that is unsupported by the provider of an \all.
For example, a company can use an internal build system that they do not expose to the public.
Similar problems are related to the supported set of compilers and their versions.

As a first task, engineers may spend a nontrivial amount of time (days, weeks) to build a working version of the \all.
That version needs to work with the compiler, compiler switches used to develop a particular application, and a specific build system.
In some cases, the support for a target \os\ itself may be experimental.
As a result, engineers themselves are responsible for porting the \all\ to a new platform.
For example, as of April 2022, \emph{jemalloc} is still not officially supported for a popular platform like i{OS}.\footnote{\url{https://github.com/jemalloc/jemalloc/issues/2086}}

\subsubsection{Version updates}

Like any other dependency exposed in the form of source code, periodic updates to the dependency's source code may need to be applied.
Typical reasons for integrating new versions include the availability of desired features, fixes for critical defects, or performance improvements.
Engineers will need to allocate some resources for periodic updates (e.g., weekly, monthly, or release basis) or continuously monitor the dependency’s code base.
Each update may again have problems described in~\Cref{subsec:build}.
For example, new code changes in \all\ may use language features that are not supported by the compiler used to build the application consuming the \all.

\subsection{Quality}

\subsubsection{Defect management}

Commercial \os s, such as Apple's Darwin or Microsoft Windows, are developed and tested by a relatively large team of engineers~\cite{lucovsky_2000, build_master_2005} compared to \oss.
The commercial \all 's code benefit from going through an extensive testing process to ensure that all the applications deployed with the \os\ are functioning correctly.
If a user discovers critical bugs after the \os 's release, there are usually channels to request fixes.
For example, a premium support contract or a  service-level agreement may guarantee a particular defect resolution time.

For systems that demand reliability, like stock trading infrastructure, having this level of access to support is critical.
In the case of custom \all s released as \oss, such formal agreements do not exist.
Projects like \emph{jemalloc} or \emph{mimalloc} have only a handful of core contributors.
There are no guarantees that any discovered issue will get a resolution during a specific timeframe.
The support channel may consist of just filing an issue on GitHub with the hope that someone will respond to it and provide an actionable resolution.

\subsubsection{Toolchain support}

During their daily work, engineers use developer productivity tools such as debuggers and profilers.
These tools help to analyze the application's \mem\ usage and debug issues related to \mem\ management.
Custom \all s
use various techniques to replace the default \all.
Those techniques include the usage of
callbacks~\cite[p. 977]{singh_mac_2016},
hooks~\cite{malloc_hooks}, or
preloading the shim~\cite{kobayashi_tips_2013}.
The productivity tools may also need to replace a default \all\ to track the \mem\ usage.
The intercept mechanism may be identical that used by a custom \all.
No support exists for more than one entity to intercept allocations in parallel.
As a result, the productivity tools are not usable with a custom \all, thus reducing the techniques and tools engineers have at their disposal to detect and fix defects.

We have witnessed anecdotal instances of productivity tools assuming the presence of a default \all.
In those cases, the tools may rely on undocumented internal implementation details such as the presence of certain linker symbols, patterns signifying how the invalid \mem\ is identified, or the location of log files.

Using a custom \all\ breaks those assumptions.
If engineers observe a \mem\ corruption or a leak, they cannot use productivity tools with the custom \memall.
Efficient debugging and profiling are possible only with the default \all.
This limitation makes it challenging to determine if the root cause of the incorrect behavior is the application, the \memall\ itself, or a combination of both.
Limited reproducibility means that engineers may first have to validate all memory-related defects with a default \all.

\subsubsection{Defect exposure}

There exist cases where the \all\ may hide certain types of errors from users or try to avoid crashing or corrupting the process heap in case of invalid input.
For example, in {C++}, it is recommended to use either operator \texttt{new} or \texttt{std::make\_unique} to allocate the storage for an object.
The expectation is that if the caller allocates \mem\ via operator \texttt{new}, the caller will use operator \texttt{delete} to release it.
However, a caller could use the \texttt{free()} function instead.
A default \all\ may allow releasing \mem\ allocated using a \say{mismatched function} but the custom \all\ will not.
Such differences in behavior may expose genuine defects in the application.

Another category of problems is related to unwritten contracts when allocating \mem.
For example, some \all s may block the call to \texttt{malloc()} until there is available \mem, and some may not.
Some \all s may subscribe to the low-\mem\ notifications from the \os\ to preemptively either release the pre-allocated \mem\ or perform an internal \mem\ compaction process.
Others may not.

\subsubsection{Support for multiple \all s}

There are various reasons why the application needs to support multiple \all s in parallel.
In our experience, the most common reasons are:

\begin{itemize}
	\item \emph{Ability to revert to the default \all}.
	A custom \memall\ may contain defects that cannot be fixed in the time allocated.
	The engineers may have to revert the application to using a default \all\ to unblock the development process.
	Alternatively, the latest \os\ update may not be compatible with how the custom \all\ works.
	For engineers to revert the dependency usage efficiently, they must make sure that the application still builds and works with various \all s.
	That requires the introduction of an additional build flavor that adds to the daily maintenance costs.

	\item \emph{Ability to compare the performance characteristics}.
	As the application's code base evolves, its behavior changes.
	Those changes can significantly impact the application's \mem\ usage patterns.
	The acceptable performance results from weeks or months ago may no longer be optimal.
	Using a custom \all\ may result in performance degradation instead of improvements~\cite{reconsider_berger}.
	Therefore, engineers must periodically run performance and stress tests using multiple \all s.

	\item \emph{Preemptive detection of compatibility}.
	For \oss, the success of a project or a company may depend on software's adoption rate.
	Adaptors of software involved in tasks that require a significant amount of resources (e.g., database engine, image processing) may use custom \all s that the authors of the component do not support.
	If the authors cannot make the component function with a custom \all, the consumer may switch to a different product.
\end{itemize}

Additional reasons we have observed in the past include discontinuing of custom \all\ development, software licensing changes, or the release of a new \all\ that provides better performance.

\subsection{Performance}

\subsubsection{Side-effects of optimizations}

When an application requests \mem, the \all\ rounds up the allocation size to a specific value (e.g., next power-of-two) depending on the algorithm it uses.
That value may be specific to a page size of an \os\ or how a particular \all\ manages \mem\ internally.
For example, when a caller allocates \num{33} bytes on mac{OS}, the actual size of allocation as reported by \texttt{malloc\_size()} is instead \num{48} bytes.
The memory allocated is \num{31}\% larger than asked for.

A standard performance optimization technique is to customize the layout of data structures.
Data structures are aligned and combined to reach a specific size to reduce waste.
The application then explicitly makes requests rounded to a specific size.
However, using a custom \memall\ may increase the request size internally because it must account for its private metadata.
Therefore, allocations assumed to be of optimal size may use significantly more memory and increase the application's resource usage.
We have observed this problem repeatedly during commercial software development.

\subsubsection{Cost of generalization}

The \oss\ \all s are intended for multiple \os s.
Each \os\ has its \krn\ \all\ and \mem\ management strategy that is different from others.
For example,

\begin{itemize}
	\item i{OS} uses a concept called \emph{compressed \mem}~\cite{levin_ios_2017}.
	Releasing \mem\ on i{OS} can require uncompressing the affected \mem\ region into the available free \mem\ to be released.
	Ironically, freeing \mem\ on i{OS} may cause an application to run out of \mem\ and possibly be costlier than just allocating \mem.

	\item Linux \krn\ uses a concept called \emph{overcommitting}.
	It means that during the allocation, \mem\ is not reserved~\cite{love_linux_2005,bovet_understanding_2003}.
	The \mem\ is allocated only when the first write to the allocated region is made.
	On the other hand, Windows explicitly tries to avoid this situation~\cite{russinovich_windows_2012}.
\end{itemize}

Understanding the nuances of \mem\ management for each \os\ is an involved task.
Engineers responsible for developing custom \all s may not have the resources to \emph{become experts for all the \os s}.
To increase the \all 's portability, the \all s limit the interface they use to interact with the \os 's \mem\ management routines.
For example, \emph{mimalloc} makes a conscious choice to use only \texttt{mmap()} to request \mem\ from the \os~\cite{leijen2019mimalloc}.

Authors of an  \all\ can choose to optimize code paths for each \os\ separately.
The primary programming language used to develop \all s is {C}.
Supporting code specific to an \os\ means the usage of many \texttt{\#ifdef} statements across the code base.
Ensuring that the \all\ builds without errors may require a different build for each preprocessor directive specific to an \os.
That implies increased costs for software maintenance.

\section{Conclusions and recommendations}

Existing research on \all s enumerate either the technical nuances necessary to produce one or the benefits of utilizing a custom \all.
This paper documents several complications that result from the usage of custom \all s.
Those issues can increase the maintenance cost of the software, expose existing bugs, and result in inadvertent performance regressions.
Most of the maintenance challenges we enumerate are also relevant to software components other than \all s.
We recommend that a project using a custom \all:

\begin{enumerate}
	\item \emph{Invest in developing the in-house expertise regarding a particular \all}.
	Engineers may need that knowledge to fix the defects in the \all 's code, tune its performance, and efficiently upgrade the source code with each new release.
	Both the initial and ongoing maintenance investments are necessary to integrate a custom memory \all\ successfully

	\item \emph{Maintain working builds of the application with both the default and custom \all s}.
	A project may have to be ready to switch back to a default \all\ at a moment's notice if defects cannot be fixed immediately or the custom \all\ causes significant performance regressions.
	Engineers may have to use the application built with the default \all\ to use specific tools such as \mem\ leak detectors or profilers.

	\item \emph{Cultivate a good relationship with the community and authors of the \all}.
	If using the custom \memall\ is critical to the product's success, then making a conscious investment into establishing good relations with the community producing the \all\ is important.
	Engineers should report all the defects found, try to improve the documentation, and contribute back any improvements made to the \all 's code base.
\end{enumerate}

\bibliographystyle{IEEEtran}
\bibliography{custom-allocators}

\end{document}